\documentclass[conference]{IEEEtran}
\IEEEoverridecommandlockouts
\usepackage{cite}
\usepackage{amsmath,amssymb,amsfonts}
\usepackage{algorithmic}
\usepackage{graphicx}
\usepackage{soul}
\usepackage{comment}
\usepackage{geometry}
\usepackage{url}
\usepackage{textcomp}
\usepackage{xcolor}
\def\BibTeX{{\rm B\kern-.05em{\sc i\kern-.025em b}\kern-.08em
    T\kern-.1667em\lower.7ex\hbox{E}\kern-.125emX}}
\begin{document}

\title{ABI: A tightly integrated, unified, sparsity-aware, reconfigurable, compute near-register file/cache GPU architecture with light-weight softmax for deep learning, linear algebra, and Ising compute}


\author{
Siddhartha Raman Sundara Raman, Jaydeep P. Kulkarni \\
The Department of Electrical and Computer Engineering, The University of Texas at Austin, Austin, Texas, 78712 \\
s.siddhartharaman@utexas.edu
}

\maketitle

\begin{abstract}
We present a tightly integrated, unified near-memory GPU architecture, delivering 6–16× speedups and 6–13× energy savings across Convolutional Neural Networks, Graph Convolutional Networks, Linear Programming, Large Language models, and Ising workloads over MIAOW GPU. Features include a custom sparsity-aware near-memory circuit ($\sim$1.5× energy savings), a custom lightweight softmax circuit ($\sim$1.6× energy savings), reconfigurable compute up to INT16 with dynamic resolution update, and scalable across problem sizes.  ABI enabled MI300/Blackwell yields $\sim$4.5× speedup over baseline MI300/Blackwell

\end{abstract}


\section{Introduction}
\begin{figure}[t]
\centering
\includegraphics[width=\linewidth]{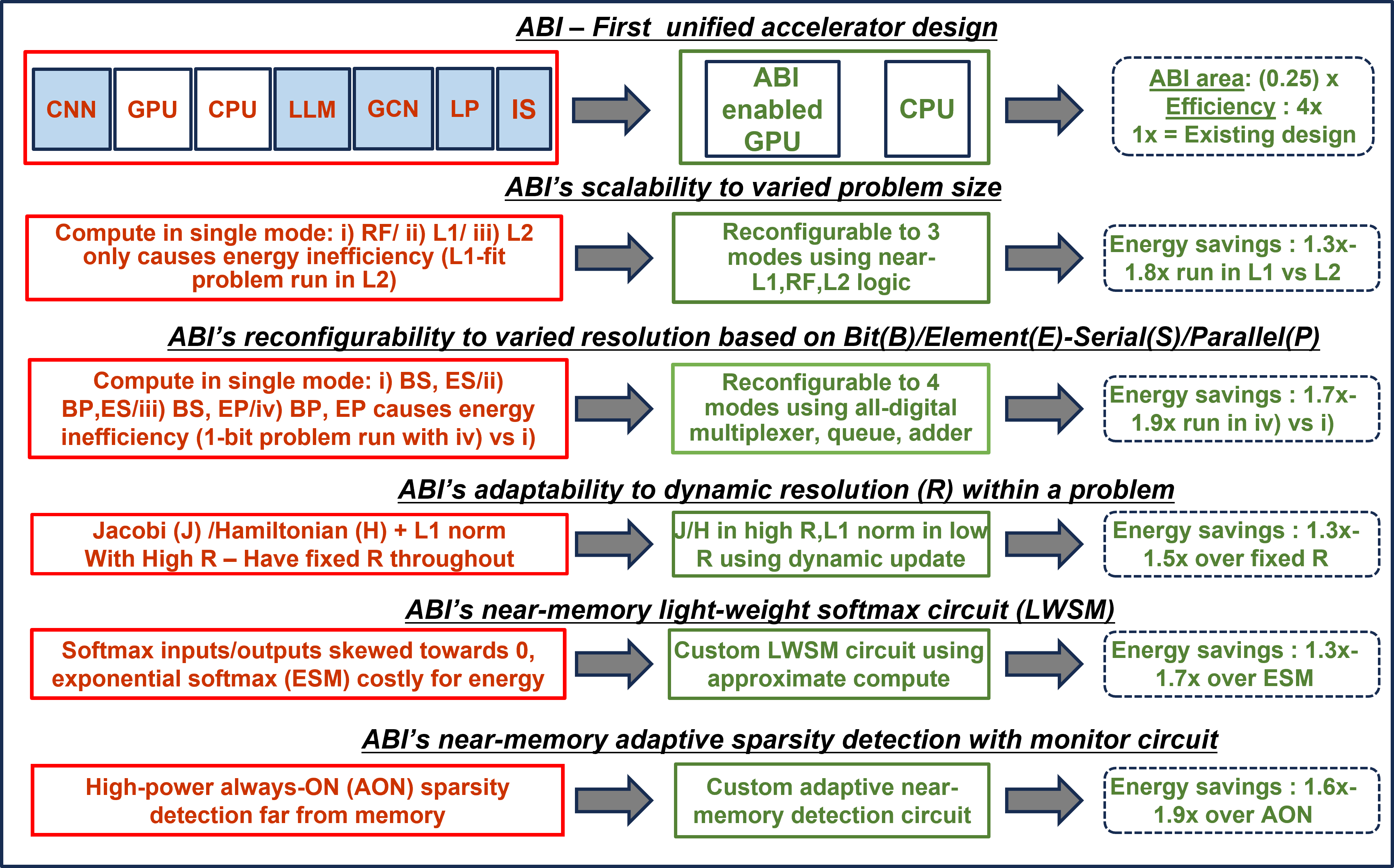}
\vspace{-2em}
\caption{\textbf{Limitations of existing accelerators (red/first column), proposed design changes to realize ABI (green/second column), resultant energy savings, area, efficiency using ABI (green/third column)  }}
\label{Fig1}
\vspace{-1.5em}
\end{figure}

Deep-learning using convolutional neural networks (CNNs), graph-processing with graph convolutional networks (GCNs), optimization via linear programming (LP) algorithms, physics-inspired computing with Ising models, and large-language models (LLMs) using transformers are among the most compute and data-intensive algorithms. To accelerate them, multiple dedicated accelerators using traditional memory technologies \cite{b1}\cite{b2}\cite{b3}\cite{b4}\cite{b10}\cite{b15}\cite{Ising_arxiv}\cite{b19}\cite{b20}\cite{NEM_GNN_arxiv}\cite{SPARK_arxiv}\cite{DRAM_PIM_arxiv} and using emerging non-volatile memory technologies \cite{b16}\cite{b17}\cite{b18} have been proposed. 
However, integrating these dedicated accelerators into congested SoCs remains area, power-intensive, unscalable, as detailed in Fig.\ref{Fig1}a). To address this, we transform existing SoCs comprising of CPU, GPU and accelerators into i) \textbf{tightly integrated design} adding minimal logic to GPU, to realize ABI enabled GPU. ii) \textbf{unified design} accelerating multiple applications with low area, and high energy efficiency (Fig.\ref{Fig1}a).  Furthermore, ABI builds on GPU’s high-throughput strengths (Multiply and Accumulate operations) but addresses its key bottleneck of frequent memory-ALU data movement, via a \textbf{compute near-memory} architecture. By leveraging high memory bandwidth and low-frequency in GPU, ABI supports near-register file/near-cache logic to efficiently accelerate diverse workloads. 

\par Building a unified design is challenging due to diverse application needs. We summarize these challenges, the limits of existing accelerators, and how ABI’s \textbf{reconfigurable approaches (R1–R3)} address them. \textbf{R1)} Application problem sizes vary widely (e.g., 10-1000 constraints in LP or 100-1M spins in Ising). Existing accelerators use only register file (RF), L1, or L2 cache for in/near-memory, causing energy inefficiencies. For instance, L1-fit problem run in L2 incurs 1.4x more energy, due to data transfers between ALU and L1/L2. ABI reconfigures near-RF/L1/L2 to match problem size. \textbf{R2)}  Problem resolutions vary within each application (e.g., 1-16 bits in Ising/CNN), so the optimal Bit-Serial/Bit-Parallel (BS/BP) and Element-Serial/Element-Parallel (ES/EP) mode depends on the problem, not the application. Fixed mode pairs in existing accelerators (e.g., BP,ES) create inefficiencies—e.g., 1-bit compute in BP,ES consumes 1.7$\times$ more energy than BS,ES. ABI adapts near-memory compute to the problem's resolution for higher energy efficiency.
\textbf{R3)} Different algorithm stages need different compute resolutions (e.g., L1 norm can use lower resolution than spin update/Jacobi in Ising/LP). ABI dynamically adjusts resolution via programmable registers, unlike fixed-resolution accelerators, achieving up to 1.25× power savings with minimal impact on solution time (Fig.\ref{Fig1}c).  
\par Exponential based softmax is area/power-heavy, and is deterrent to near-memory compute. We propose an approximate compute block that leverages custom \textbf{lightweight near-memory softmax} (LWSM) circuit to enable area/energy-efficient near-memory softmax compute with 1.5× energy savings and $\sim$99\% accuracy (Fig.\ref{Fig1}d).
\par Algorithms have varying sparsity, and lead to high power in sparsity detection logic, if detection is not adaptive. i.e. always-on sparsity detection logic incurs power in low-sparsity cases. ABI uses a i) programmable sparsity monitor that disables detection units when unused. ii) To fit sparsity detection near-memory (Fig.\ref{Fig1}e), ABI adds a custom \textbf{sparsity-aware} circuit to achieve $\sim$1.8x energy savings.
\par This article presents the first tightly integrated, sparsity-aware, reconfigurable near-RF/cache GPU with LWSM, supporting multiple applications. Section II details the overall design; Section III reconfigurable architecture; Section IV LWSM; Section V sparsity handling; Section VI unified architecture with TSMC 65nm test-chip results.

\begin{figure}[t]
\centering
\includegraphics[width=\linewidth]{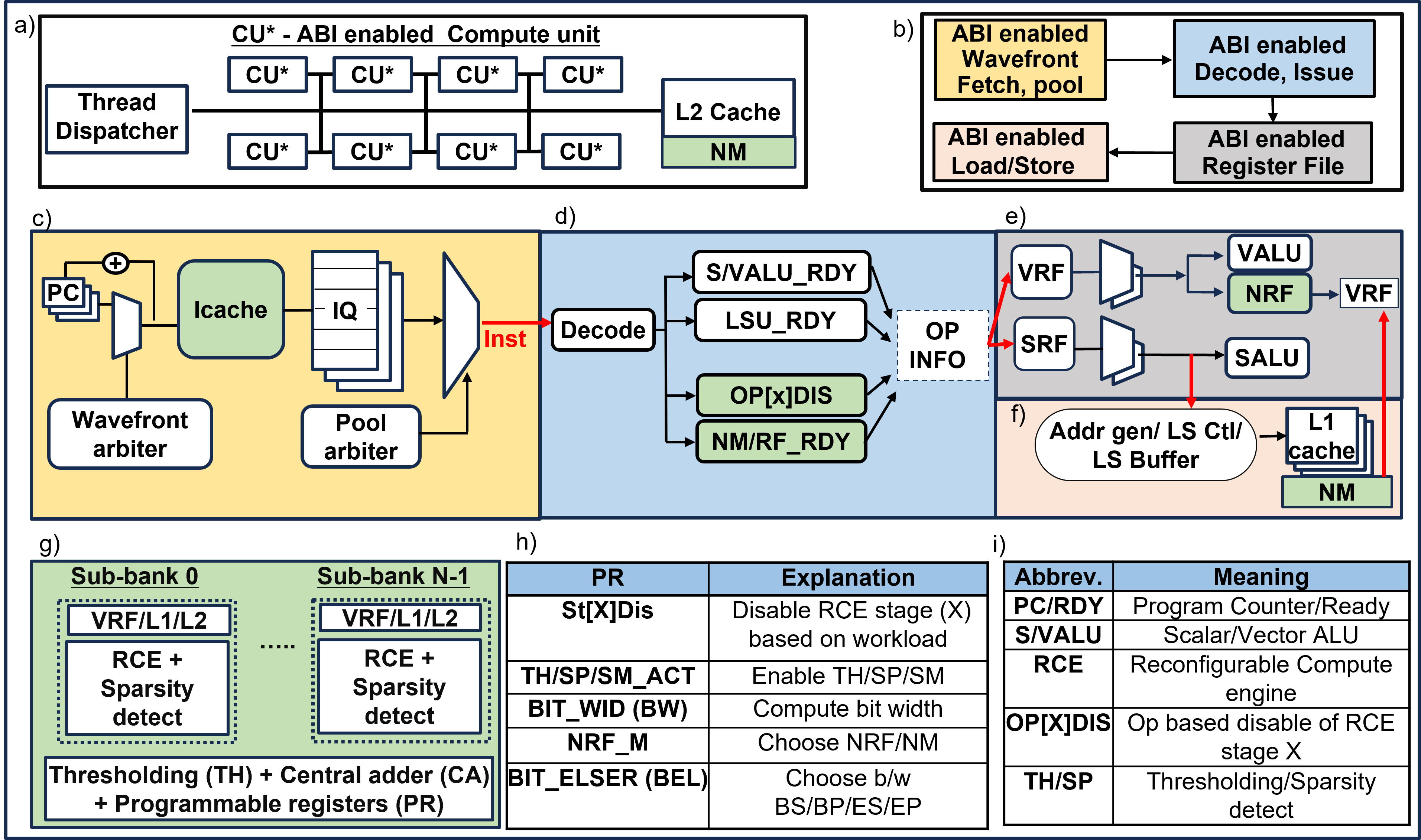}
\vspace{-2em}
\caption{\textbf{ABI enabled a) \ul{tightly integrated} GPU including dispatcher, compute unit, L2 cache, b) compute unit c) wavefront fetch, pool d) decode, issue e) register file f) load/store units. g) Near-memory(NM) / Near-RF(NRF) logic floorplan h) Programmable registers i) Legend  }}
\label{Fig2}
\vspace{-1em}
\end{figure}

\section{Tightly integrated near-register file/cache design}
Fig.~\ref{Fig2} shows the ABI-enabled GPU tightly integrated using the open-source MIAOW design~\cite{b12}, programmed using Southern Islands ISA. The baseline GPU includes compute units (CUs), L2 cache, a thread dispatcher connected to the host CPU and DRAM, and closely resembles industry GPUs~\cite{b12} (Fig.~\ref{Fig2}a). Each CU has wavefront fetch/pool, decode and issue, scalar/vector register files (S/VRF), ALUs, and load-store units (LS - queues/L1 cache) (Fig.\ref{Fig2}b). ABI incorporates near-register file logic (NRF), near-L1 logic (NM) in all CUs, and near-L2 (NM) logic. The wavefront fetch/pool is unchanged and is responsible for mapping wavefronts into different CUs (Fig.\ref{Fig2}c), while decode and issue block now includes logic to handle new instructions for programming NM/NRF logic and instruction dependent disables (Fig.\ref{Fig2}d). In ABI, the S/VRF sub-banks, which in the baseline GPU broadcast data to execution units on the read, now include NRF logic for execution (Fig.\ref{Fig2}e). ABI adds data paths from NRF/NM logic in addition to existing ALU writes. The LS unit incorporates near-L1 NM logic absent in the baseline (Fig.~\ref{Fig2}f). NM/NRF logic (Fig.~\ref{Fig2}g) includes the reconfigurable compute engine (RCE), per-bank sparsity detection, thresholding, scaler (S), central adder (CA), and shared programmable registers (PRs), enabling MAC and thresholding operations. Different PRs (Fig.~\ref{Fig2}h) selectively disable RCE stages and configure sparsity (\texttt{SP\_ACT}), thresholding (\texttt{TH\_ACT}), softmax (\texttt{SM\_ACT}), memory level (\texttt{NRF\_M}), near-memory mode (\texttt{BIT\_ELSER}), and resolution (\texttt{BIT\_WID}).

\begin{figure}[t]
\centering
\includegraphics[width=\linewidth]{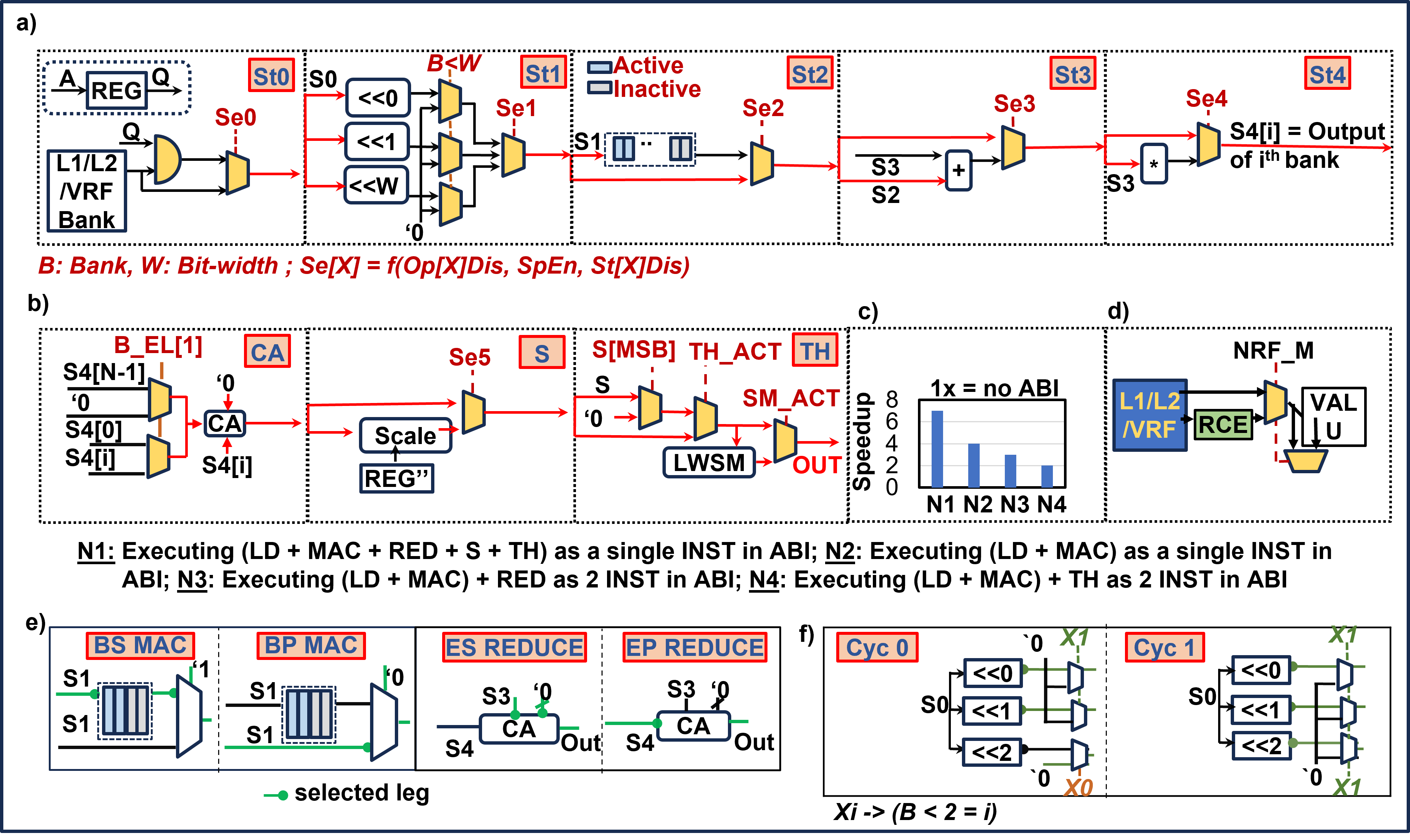}
\vspace{-2em}
\caption{\textbf{a) \ul{Reconfigurable} Compute Engine (RCE) with b) thresholding, central adder, scaler c) Speedup showing importance of RCE. ABI’s \ul{reconfigurability} circuit for d) NM/NRF compute e) varied resolution f) dynamic resolution update}}
\label{Fig3}
\vspace{-1em}
\end{figure}

\section{Reconfigurable architecture}

Fig.\ref{Fig3} shows RCE’s 5-stage unified architecture (Fig.\ref{Fig3}a).
St0: Generates partial/bit-wise dot-products via AND of L1/L2/RF reads with REG.
St1: Shifts these dot-products for multi-resolution support.
St2: Performs bit-serial accumulation using active registers.
St3: Further accumulates St2 outputs for bit-serial processing.
St4: Multiplies St3’s result with REG’’ for element-serial processing. Control logic configures stages per application and disables unused ones (e.g., Ising disables St2/St4 since spins are single-bit with no final multiply, while GCN/LLM use all stages for MAC and adjacency operations). Outputs flow through the CA/scaler and softmax/thresholding blocks (Fig.~\ref{Fig3}b). The thresholding block supports ReLU (\texttt{TH\_ACT}=1), comparison/L1 norm (\texttt{TH\_ACT}=0, \texttt{SM\_ACT}=0), and includes a lightweight near-memory softmax (LWSM). ABI fuses load, MAC, reduction, and thresholding into a single operation, reducing instructions. ABI completes VMAC/VRED in 2 cycles with NRF and 4-10 cycles with NM, enabling 2-7$\times$ speedup (Fig.~\ref{Fig3}c).

\par \textbf{R1)} Scalability to varying problem sizes is enabled by programming \texttt{NRF\_M} to select NRF or NM compute (Fig.~\ref{Fig3}d). \textbf{R2)} Resolution reconfigurability uses BS/BP and ES/EP modes by programming \texttt{BIT\_ELSER}, setting \texttt{DIS\_STAGE2}, and adjusting the CA. BS serializes 4-bit partial dot-products via one active St2 register, while BP bypasses St2. For ES, CA reduces one bank at once, with rest forced to zero (Fig.~\ref{Fig3}e); for EP, CA reduces all banks simultaneously. \textbf{R3)} Dynamic resolution is supported by programming \texttt{BIT\_WID} (up to INT16) to control St1 and mask shifter outputs. For 2-bit compute, the shift-2 path is masked; while for 3-bit compute it is enabled (Fig.~\ref{Fig3}f).
\section{Lightweight Softmax compute}
Fig.\ref{Fig4}a) illustrates ABI’s light-weight softmax (LWSM) approach, housed inside TH block, and relying on approximate compute. Firstly, the approximate computing block performs (1+x) (marked as In1)/ $\sum$(1+x) (marked as UpdCnt), by exploiting x being near 1. Instead of division, it finds the position of first ‘1’ in In1 and in UpdCnt.  The difference between the 2 positions, when shifted/ decoded yields softmax output. Second, a custom circuit maps the fixed dataflow of summation, find-first, and difference (Fig.~\ref{Fig4}a). It i) exploits the fact that one of the inputs in 3-input addition is likely ‘1’, and can be reused between performing summation, difference, thereby reducing area. ii) uses an LSB-to-MSB find-first search, since the first ‘1’ typically appears near the LSBs, limiting the search range and further reducing gate depth. Approximation of softmax from division to difference,shifter delivers 2$\times$ area savings, 1.6$\times$ speedup, and $<0.1\%$ accuracy loss by reducing gate depth, easing routing near memory, and improving timing versus a standard-cell synthesized design, enabling efficient near-memory softmax compute.

\section{Sparsity awareness}
\begin{figure}[t]
\centering
\includegraphics[width=\linewidth]{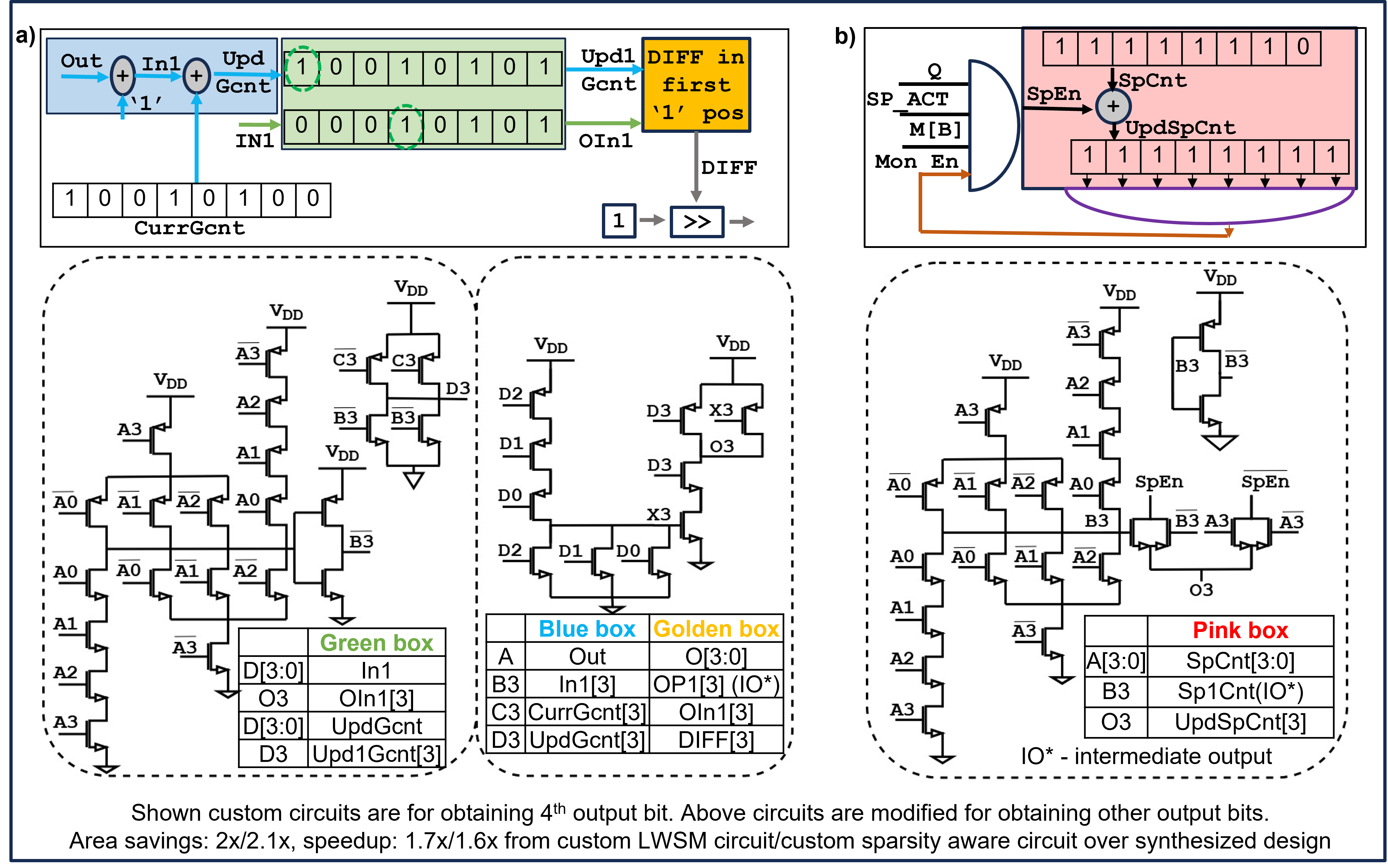}
\vspace{-2em}
\caption{\textbf{Approximate compute block diagram, custom circuit, mappings between block diagram and custom circuit for a) \ul{light-weight softmax} (LWSM) b) \ul{Sparsity} monitor necessary for low area, high speed near-memory compute}}
\label{Fig4}
\vspace{-1em}
\end{figure}

 Adaptive sparsity-awareness uses the near-memory sparsity-detect dataflow in Fig.~\ref{Fig4}b, which monitors RF/L1/L2 reads and REG. If any are zero, SpEn activates gate St1-3 of RCE, saving power. The sparsity monitor turns on SpEn for 512 (programmable up to $2^{16}$ consecutive cycles. If SpEn remains off, the monitor preemptively shuts down the sparsity detection block and \texttt{SP\_ACT} to save power. This fixed dataflow is mapped onto a custom circuit using transmission gate that exploits the fact that UpdSpCnt can be SpCnt+1 or SpCnt, with a greater possibility of SpCnt+1. Thus, SpCnt+1 is precomputed and early-gated to save power, enabling transmission-gate multiplexing with lower fanout, cutting area by 2$\times$ and improving speed by 1.6$\times$ over a synthesized AndOrInvert(AOI)-based design \cite{b27}\cite{b11}\cite{Cryo_arxiv}.

\section{Results}
\subsection{Test chip/measurement setup}
\begin{figure}[t]
\centering
\includegraphics[width=\linewidth]{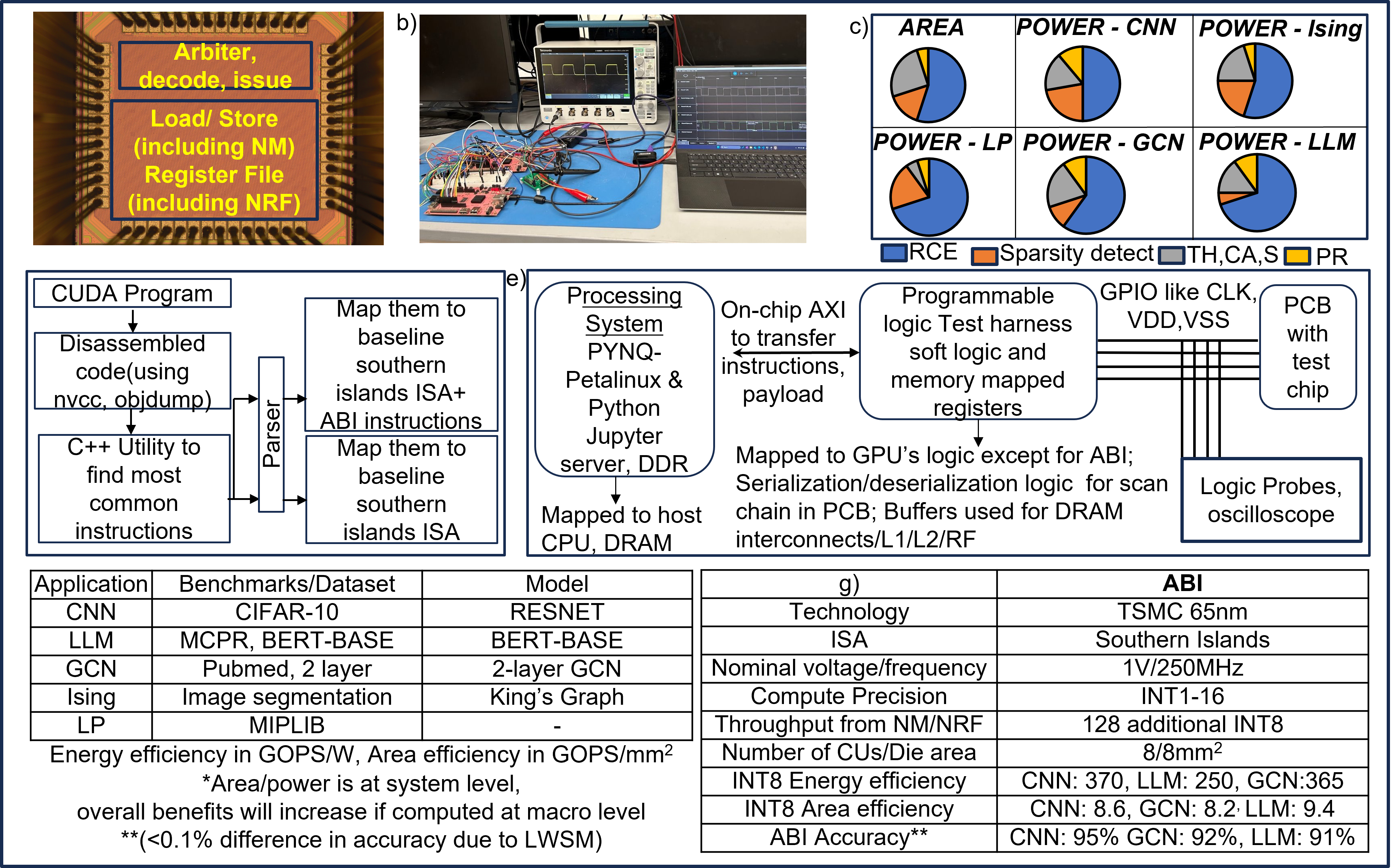}
\vspace{-2em}
\caption{\textbf{a) Die photograph b) Measurement setup c) Area, power breakdown for CNN, LP, GCN, Ising, LLM across RCE, sparsity, TH, CA, S, PR d) Offline program flow e) Programming model f) Benchmarks g) Parameters of ABI }}
\label{Fig6}
\vspace{-1em}
\end{figure}

Fig.\ref{Fig6}a) shows the die-photograph of chip, highlighting wavefront arbiter, decode, issue,  load store unit, NM/NRF logic, RF. Fig.\ref{Fig6}b) shows the measurement setup, Fig.\ref{Fig6}c) highlights the area/power distribution different applications. Fig.~\ref{Fig6}d) outlines the offline flow for programming the ABI-enabled GPU to obtain performance and energy results for CNN, Ising, LP, GCN, and LLM workloads. CUDA kernels are written, disassembled with \texttt{nvcc}/\texttt{objdump}, and the most common instructions are extracted, mapped to Southern Islands ISA for execution on ABI-enabled GPU. These instructions also program OP[X]\_DIS to disable selected near-memory and near-register-file stages. Fig.~\ref{Fig6}e) shows the ZYNQ-7000 FPGA workflow used to measure the ABI-enabled GPU. PYNQ-Petalinux and Jupyter combination send inputs and instructions to the GPU via on-chip AXI, to program ABI. Serialized inputs go to the test chip, and outputs from ABI’s near-memory logic are scanned, probed by oscilloscopes, deserialized, and monitored by the processing system Fig.\ref{Fig6}f) 

\subsection{Unified architecture}
\begin{figure}[t]
\centering
\includegraphics[width=\linewidth]{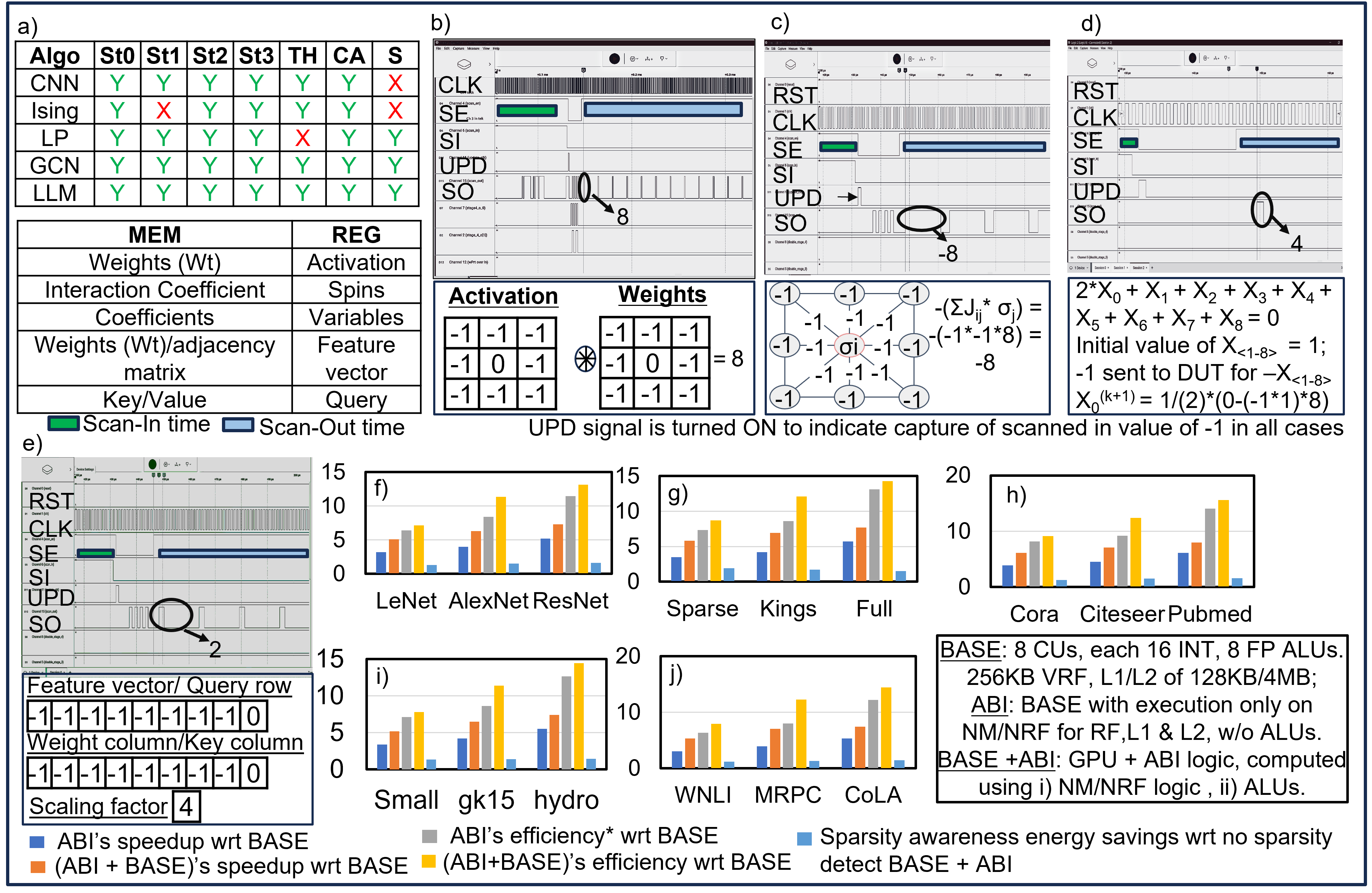}
\vspace{-2em}
\caption{\textbf{a) Hardware mapping for \ul{unified architecture}, b) Oscilloscope capture with output values circled for b) CNN c) Ising d) LP e) GCN/LLM. ABI, (ABI+BASE) Speedup, Energy efficiency, energy savings from sparsity awareness for f) CNN g) Ising h) GCN i) LP j) LLM wrt BASE}}
\label{Fig5}
\vspace{-1em}
\end{figure}

 Oscilloscope captures (Fig.\ref{Fig5}b–e) validate NRF functionality across different applications, using identical inputs, while mapping onto ABI differently (Fig.\ref{Fig5}a) but producing workload-specific outputs.

\par \underline{CNN}: Weights stay in memory (weight-stationary), while activations reside in registers. St0–St3 compute partial dot-products for convolution and linear layers. CA accumulates bank outputs; scaling (S) is disabled, and TH applies ReLU when needed. LWSM performs final label selection. Fig.~\ref{Fig5}b shows NRF compute for a 3×3 CNN convolution: activations are loaded into REG, scanned into the DUT, and the update clock programs the logic. Each bank computes partial products (e.g., $(-1)\cdot(-1)$), and CA accumulates them to produce 8.
\par \underline{Ising}: Interaction coefficients reside in memory and spins in REG, following the IC-stationary design of \cite{b3}. St0–St4 iteratively compute $H_{\sigma}$ (Fig.~\ref{Fig5}c), with St0–St3 evaluating $J_{ij}\sigma_j$ and CA summing across banks to produce $H_{\sigma}=\sum J_{ij}\sigma_j$. TH compares $H_{\sigma}$ to 0 and is also used for the L1 norm. St1 is disabled since spins are single-bit, and S/LWSM are unused. Fig.~\ref{Fig5}d shows NRF capture for $H_{\sigma}$ on a King’s graph (8 neighbors): spins \(-1\) are scanned into REG, ICs \(-1\) stored in memory, and CA accumulates to \(-8\), which is scanned out.
\par \underline{LP}: Constraint coefficients $(C,D)$ stay in memory and variables $X$ in REG, following a coefficient-stationary design. St0–St3 compute $(b_i - a_{ij}x_j^{(k)})$, with S applying the scale $1/a_{ii}$. CA accumulates $x_i^{(k+1)}$ and performs subtraction, keeping TH and LWSM gated off. Fig.~\ref{Fig5}d shows NRF capture for updating $X_0$ with eight variables initialized to $-1$ and all coefficients set to $1$ except for $X_0$. REG$''$ holds $2$, while $C_{ij}$ and $X_j=-1$ load into memory and REG. Accumulating $(-1)\cdot(-1)$ across banks gives $8$, which S scales by $2$ to produce $4$.
\par \underline{GCN}: Weights and the adjacency matrix reside in memory, with the feature vector in REG (weight-stationary). All RCE stages, CA, TH, and S are enabled for combination and aggregation, with bank parallelism supporting simultaneous compute of both. For combination, St0–St3 compute dot products, CA reduces bank outputs, S scales by neighbor count, and TH applies softmax. For aggregation, the combination result is written to REG, multiplied with the adjacency matrix via St0–St3, and reduced by CA. Fig.~\ref{Fig5}e shows NRF capture for combining a $1\times8$ feature vector with an $8\times1$ weight column (4-neighbors): each bank's output is summed, scaled and yields 2.
\par \underline{LLM}: Key and Value matrices reside in memory, and the Query matrix is stored in REG. As in GCN, all RCE stages plus TH, S, and CA are enabled. The GCN combination step corresponds to the Q–K multiplication in LLMs: St0–St3 compute the Q·K dot product, S scales by the embedding count, and TH applies softmax. Aggregation mirrors multiplication with the Value matrix (softmax ignored here). Figure~\ref{Fig5}e shows the NRF compute for Q·K: St0–St3 compute the dot product, CA accumulates across banks, and S scales by 4 (stored in REG''), yielding an output of 2.

\subsection{Comparison results}
 We compare: (i) baseline MIAOW GPU (BASE), (ii) ABI (MIAOW with execution only on NM/NRF logic), and (iii) BASE+ABI (ALUs plus NM/NRF) in Fig.\ref{Fig5}f-j. ABI delivers 3-6x speedup over BASE through single-operation execution, reduced data movement, and higher throughput. BASE+ABI achieves 6-16$\times$ speedup by running ALUs in parallel with NM/NRF. Efficiency improves 5-13$\times$ due to lower power from reduced data movement, minimal queue use, fully gated ALUs (in ABI), sparsity-aware gating (1.5$\times$ savings), and NM logic leveraging RF/L1/L2 throughput. In BASE+ABI, ALU throughput offsets added power, giving up to 15x improvement. INT2 is efficient than INT8 due to lower shifter/register power and more operations per cycle from reduced precision.

\begin{figure}[t]
\centering
\includegraphics[width=\linewidth]{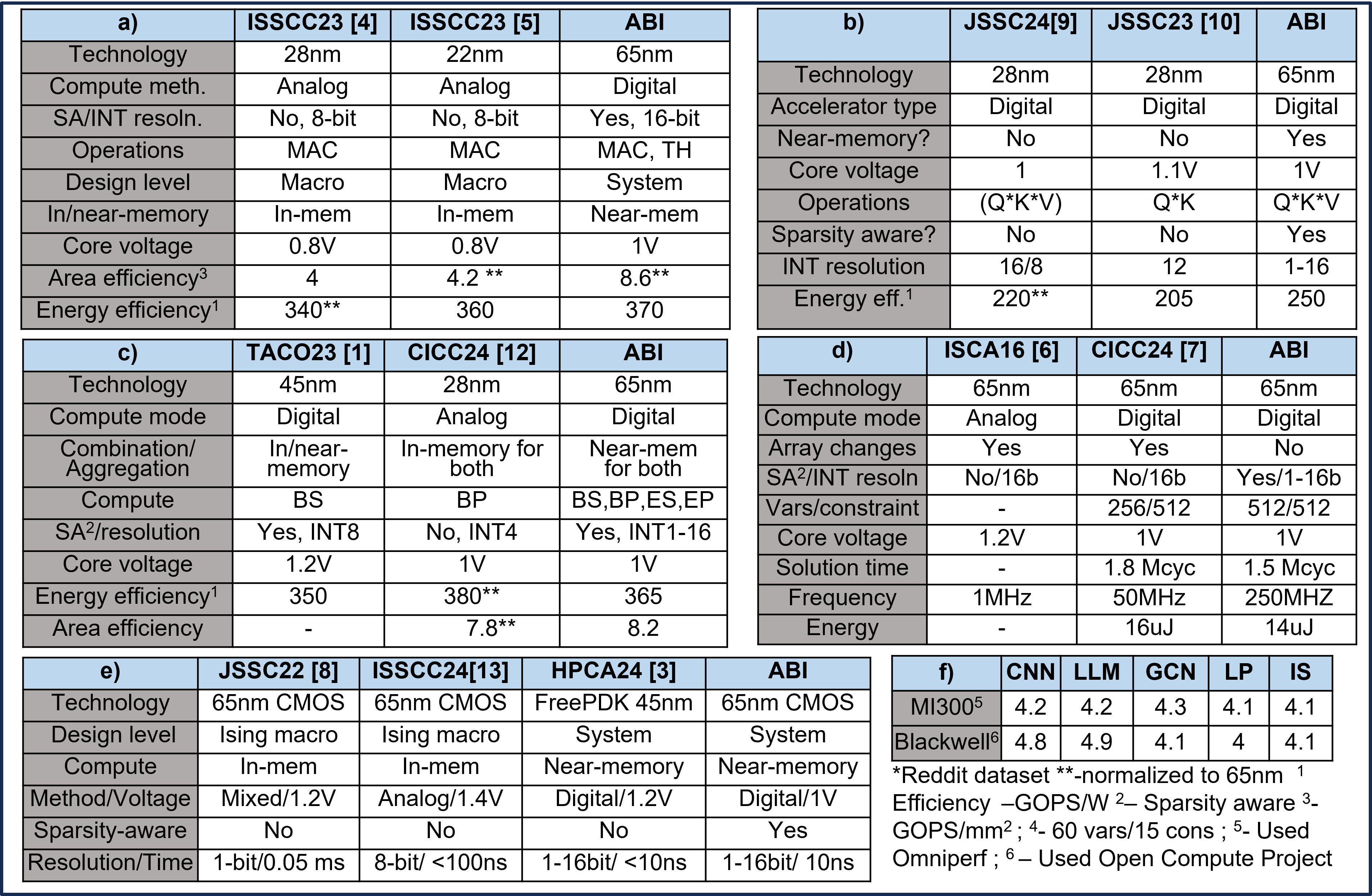}
\vspace{-2em}
\caption{\textbf{Comparison among existing dedicated accelerators for a) CNN b) LLM c) GCN d) LP e) Ising f) Estimated speedup (using Omniperf, Open compute project) of MI300/Blackwell if ABI’s NM/NRF logic is embedded onto them}}
\label{Fig7}
\vspace{-1em}
\end{figure}

Fig.\ref{Fig7}a-e) compares against the existing dedicated accelerators, While recent accelerators are analog, lack sparsity awareness, and are optimized solely for MAC operations, ABI is fully digital, supports both MAC and thresholding, and operates near-memory, with sparsity awareness at 1V. ABI enabled GPU achieves higher area efficiency by reusing GPU resources with minimal near-memory logic. Its superior energy efficiency stems from increased throughput via RF/L1/L2, sparsity awareness, reduced instruction execution, LWS, and integrated thresholding. We observe that ABI achieves competitive energy efficiency of $\sim$370 GOPS/W, operating at a frequency of 250MHz. If existing NVIDIA/AMD GPUs (Fig.\ref{Fig7}f) are embedded with ABI’s near-memory architecture, the estimated speedup is $\sim$4x.
\section{Conclusion}
We present the first unified, sparsity-aware design that integrates reconfigurable near-memory compute into a GPU for CNNs, Ising compute, LPs, transformers, and GCN in TSMC65nm. We achieve speedups of 6-16x and energy savings of 6-13x over MIAOW GPU. Furthermore, a custom sparsity-aware near-memory circuit enables 1.5x savings, a custom LWSM circuit helps achieve 1.6x energy savings, with energy-efficiency of 370GOPS/W, at frequency of 250MHz.  

\begin{figure}[t]
\centering
\includegraphics[width=\linewidth]{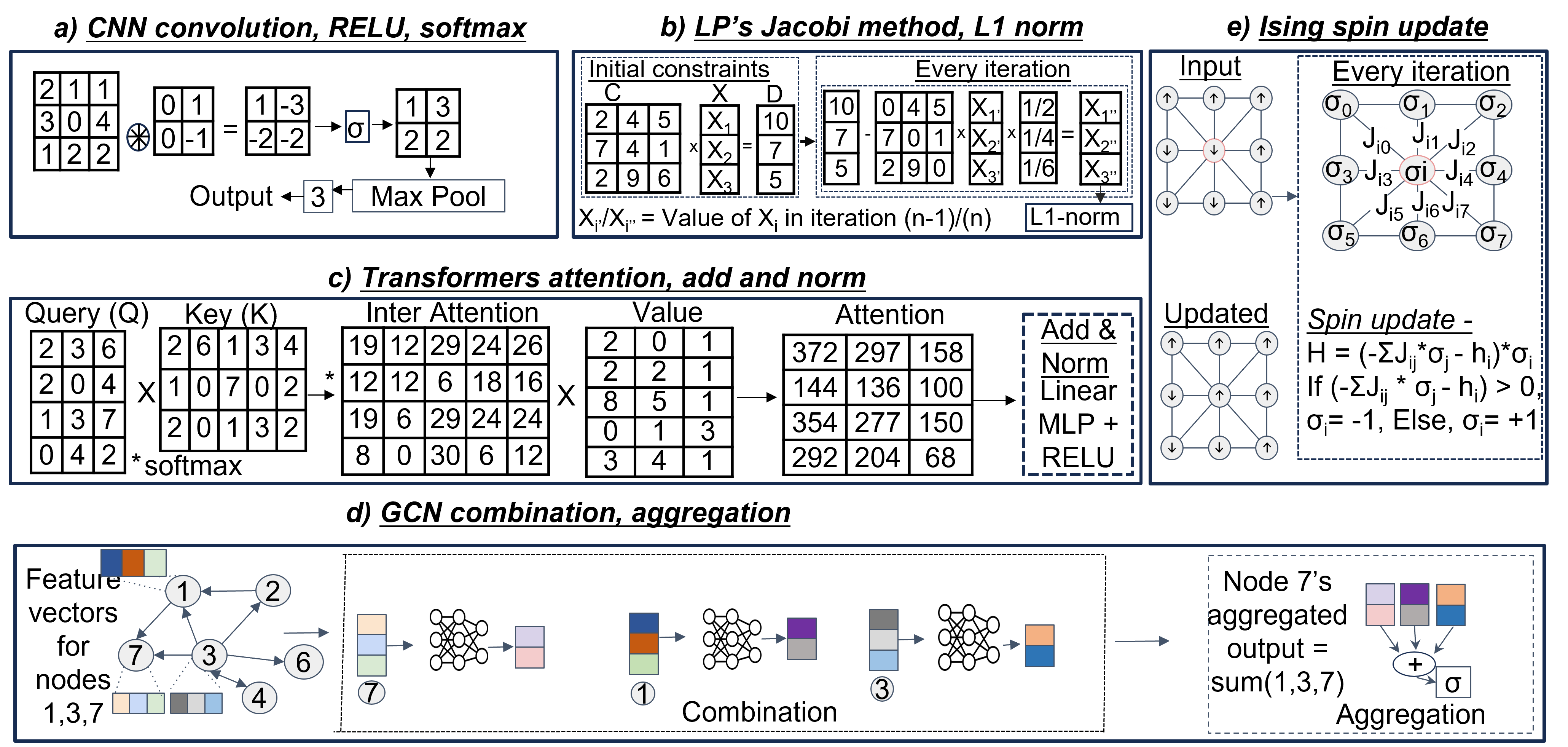}
\vspace{-2em}
\caption{Examples of \textbf{a) CNN convolution, b) Linear programming c) Transfomer engine, attention, add and norm d) GCN combination, aggregation e) Ising compute }}
\label{Algos}
\vspace{-1em}
\end{figure}

\section{Appendix}
\subsection{Bitcell design}
This architecture shown here currently uses Static Random Access Memory (SRAM)/flip-flop based implementation, however, it is not necessary that this needs to be SRAM/flip-flop. There are no assumptions made on the underlying bitcell design, and is extendable to different bitcell design including emerging non-volatile memory technologies \cite{b21} like Resistive Random Access Memory (RRAM) \cite{b22} \cite{b23}, Ferroelectric Field Effect Transistors (FeFET) \cite{b24}, Magnetoresistive Random Access Memory (MRAM) \cite{b25}, Phase Transition Material (PTM) assisted designs \cite{b26}.
\subsection{Operating temperature}
We don't make any assumptions about the underlying temperature, the performance numbers are different at cryogenic temperature however, one can realize this design using any form of designs

\subsection{Algorithms}
Fig.\ref{Algos} shows examples of the algorithms that are computed using ABI, starting from convolutional neural network, linear programming , transformer engine (specifically attention, add and norm compute), graph neural networks to Ising machines.

\end{document}